\documentstyle[epsfig]{mn}
\begin{document}
\input ./BoxedEPS.tex
\SetEPSFDirectory{./}
\SetRokickiEPSFSpecial
\HideDisplacementBoxes

\title[ELAIS paper II: mid-infrared extragalactic source counts]
{The European Large Area
ISO Survey II: mid-infrared extragalactic source counts} 
\author[Stephen Serjeant, Seb Oliver, Michael Rowan-Robinson, et al.]{
Stephen Serjeant$^{1}$, Seb Oliver$^{1}$, Michael
Rowan-Robinson$^{1}$, 
Hans Crockett$^{1}$,\vspace*{0.200cm}\\ 
{\LARGE Vasileios Misoulis$^{1}$, Tim Sumner$^{1}$, Carlotta
Gruppioni$^{1,2}$, Robert G. Mann$^{1,7}$,}\vspace*{0.200cm}\\
{\LARGE 
Nick Eaton$^{1}$, David Elbaz$^{3}$, David
L. Clements$^{11,4}$,
Amanda Baker$^{3,4}$, 
} \vspace*{0.200cm}\\
{\LARGE  
Andreas Efstathiou$^{1}$, 
Catherine
Cesarsky$^{3}$, Luigi Danese$^{13}$, 
Alberto Franceschini$^{5}$, 
}\vspace*{0.200cm}\\
{\LARGE 
Reinhardt
Genzel$^{6}$,
Andy Lawrence$^{7}$, Dietrich Lemke$^{8}$, 
Richard G. McMahon$^{9}$,
}\vspace*{0.200cm}\hspace*{0cm}\\ 
{\LARGE George
Miley$^{10}$,
Jean-Loup Puget$^{11}$, Brigitte
Rocca-Volmerange$^{12}$
}\vspace*{0.200cm}\\  
\parbox{159mm}{
$^{1}$ Astrophysics Group, Blackett Laboratory, Imperial College of 
Science Technology \& Medicine, Prince Consort
Rd.,London.SW7 2BZ\\
$^{2}$ Osservatorio Astronomico di Bologna, via Ranzani 1, 
40127 Bologna, Italy\\
$^{3}$ CEA / SACLAY, 91191 Gif sur Yvette cedex, France\\
$^{4}$ Dept of Physics \& Astronomy, Cardiff University, PO Box 913,
Wales CF24 3YB\\
$^{5}$ Dipartimento di Astronomia, Universit\'{a} di Padova, Vicolo
Osservatorio 5, I-35122 Padova, Italy\\
$^{6}$ Max-Planck-Institut f\"{u}r extraterrestrische Physik, 
P.O.Box 1603, 85740 Garching, Germany\\
$^{7}$ Institute for Astronomy, University of Edinburgh, Royal
Observatory, Blackford Hill, Edinburgh EH9 3HJ\\
$^{8}$ Max-Planck-Institut f\"{u}r Astronomie, K\"{o}nigstuhl 17, D-69117, 
Heidelburg, Germany\\
$^{9}$ Institute of Astronomy, The Observatories, Madingley Road,
Cambridge, CB3 0HA\\ 
$^{10}$ Leiden University, P.O. Box 9513, 
NL-2300 RA Leiden, The Netherlands\\
$^{11}$ Institut d'Astrophysique Spatiale, 
B\^{a}timent 121, Universit\'{e} Paris XI, 91405 Orsay cedex, France\\
$^{12}$ Institut d'Astrophysique de Paris, 98bis Boulevard Arago, 
F 75014 Paris, France\\ 
$^{13}$ SISSA, International School for Advanced Studies, 
Via Beirut 2-4, 34014 Trieste, Italy
}}
\date{Accepted ;
      Received ;
      in original form 11 June 1999}
 
\pagerange{\pageref{firstpage}--\pageref{lastpage}}
\pubyear{1999}
\volume{}

\label{firstpage}

\maketitle


\begin{abstract}
We present preliminary
source counts at $6.7\mu$m and $15\mu$m from the Preliminary 
Analysis of the European Large Area ISO survey, with limiting flux
densities of $\sim2$mJy at $15\mu$m and $\sim1$mJy at $6.7\mu$m. 
We separate the
stellar contribution from the extragalactic using 
identifications with APM sources made with the likelihood ratio
technique. 
We quantify the completeness and reliability of our source extraction 
using (a) repeated
observations over small areas, (b) cross-IDs with stars of known
spectral type, (c) detections of the point spread function wings
around bright sources, (d) comparison with independent 
algorithms. 
Flux calibration at $15\mu$m was performed using stellar
identifications; the calibration does not agree with the pre-flight
estimates, probably due to effects of detector hysteresis and
photometric aperture correction. 
The $6.7\mu$m extragalactic counts are broadly
reproduced in the Pearson \& Rowan-Robinson model, but the
Franceschini et al. (1997) model underpredicts the observed source
density by 
$\sim0.5-1$ dex, though the photometry at $6.7\mu$m is still
preliminary. 
At $15\mu$m the extragalactic counts are in 
excellent agreement with the predictions of the Pearson \&
Rowan-Robinson (1996), Franceschini et al. (1994), 
Guiderdoni et al. (1997) and the evolving models of Xu et al. (1998),  
over $7$ orders of magnitude in $15\mu$m flux density. 
The counts agree with other estimates from the ISOCAM instrument at
overlapping flux densities (Elbaz et al. 1999), provided a consistent
flux calibration is used. 
Luminosity evolution at a rate of $(1+z)^3$, incorporating mid-IR
spectral features, provides a better fit to the $15\mu$m
differential counts than $(1+z)^4$ density evolution. 
No-evolution models are excluded, 
and implying that 
below around $10$ mJy at $15\mu$m the
source counts become dominated by an evolving cosmological
population of dust-shrouded starbursts and/or active
galaxies. 
\end{abstract}
\maketitle

\hspace*{0cm}
\pagebreak[4]
\section{Introduction}\label{sec:introduction}

The IRAS mission enjoyed huge successes, including the
sensational discoveries of ultra- and hyper-luminous galaxies and
of an enormous population of evolving starbursts. 
However, the survey had several drawbacks.
For example, the bright limiting flux densities restricted the 
samples to low redshifts ($z\stackrel{<}{_\sim}0.3$) for all but
a few ultraluminous objects. Also, only $\sim1000$ galaxies were
detected at $12\mu$m over the whole sky. 
These deficiencies 
restricted the study of IR-luminous
galaxies at all redshifts. 

The Infrared Space Observatory
(ISO) offered $\sim\times 1000$ improvements in sensitivity 
in the mid-IR over IRAS,
and the large allocations of guaranteed and discretionary time for
deep surveys on ISO will greatly improve on the IRAS surveys in the
mid-IR. 
For instance, ISO observations of the northern Hubble Deep
Field (Serjeant et al. 1997, 1999, Goldschimdt et al. 1997, Oliver et
al. 1997, Aussel et al. 1999, D\'{e}sert et al. 1999) reached the 
$15\mu$m confusion limit ($\sim0.1$ mJy) over $17$ square arcminutes, 
while the CAM-Deep and CAM-Shallow surveys (Elbaz et al. 1998a,b) were
slightly less sensitive 
but had wider areal coverage ($0.5$ mJy over $0.3$ 
square degrees and $0.8$ mJy over $0.41$ square degrees). 
These have also been complemented by deep ISO photometry 
of selected high-$z$ galaxies (e.g. Flores et al. 
1999).

The European Large Area ISO Survey (ELAIS, Oliver et al. 1999 (paper
I), Rowan-Robinson et al. 1998) was the largest open time
project on ISO, complementing the deep ISO samples by surveying 
$\sim12$ square degrees to a depth of $\sim2$ mJy at $15\mu$m and
$\stackrel{<}{_\sim}100$ mJy at $90\mu$m. Around half the area was also
mapped at $6.7\mu$m to $\sim1$ mJy. 
Three fields in the Northern hemisphere (N1, N2, N3) collectively
comprised around two-thirds of the $15\mu$m areal coverage, with
the remaining area taken by the Southern S1 field and several small
areas in both hemispheres. 
The ambitious cosmological aims include tracing the
extinguished star formation history of the Universe to $z\sim1-2$,
orientation-independent selection of dust-shrouded quasars, and the
potential discovery of hyperluminous galaxies (with comparable
intrinsic luminosities to IRAS FSC
10214+4724) out to redshifts $z\stackrel{<}{_\sim}5$. 
A more detailed discussion of the diverse scientific aims of ELAIS, 
the selection of areas and observational parameters can be
found in the ELAIS survey paper (Oliver et
al. 1999); in summary, 
the survey areas were selected to have low galactic
cirrus emission, high visibility by ISO, high ecliptic latitude and
avoiding $12\mu$m IRAS sources brighter than $0.6$ Jy. 
In another companion paper, Efstathiou et al. 1999, we discuss the 
$90\mu$m source counts from the Preliminary Analysis of the ELAIS
ISOPHOT data, and in Crockett et al. (1999) we discuss the stellar
mid-infrared source counts. 
The ELAIS areas have also been the subject of intensive
multi-wavelength follow up, summarised to date in Oliver et al. (1999) 
and presented in detail in other papers (e.g. Ciliegi et al. 1999,
Gruppioni et al. 1999). 
Here we
present the completeness, reliability and extragalactic source counts
from our initial Preliminary $6.7\mu$m and $15\mu$m ISOCAM catalogues. 
A future paper will present the Final Analysis products from the
ISOCAM ELAIS data, which is expected to improve on the Preliminary
Analysis presented here. 

This paper is structured as follows. In section \ref{sec:method} we
describe the Preliminary Analysis 
CAM pipeline, explaining the artefacts in the
data (section \ref{sec:method_intro}), and the pipeline algorithm (section
\ref{sec:cam_pipe}). The results from the Preliminary Analysis 
catalogue are
presented in section \ref{sec:results}. Our various completeness and
reliability 
estimates are discussed in section \ref{sec:comp_rel}, and the
segregation of 
extragalactic from stellar sources in section
\ref{sec:optid}. Section \ref{sec:counts} presents the source counts
in both wavebands. These results are compared with source count models and
previous results in section \ref{sec:discussion}, where we also
discuss the 
implications for the evolution of star forming galaxies and on the
star formation history of the Universe. 

\section{ELAIS CAM Preliminary Analysis}\label{sec:method}
\subsection{Data quality}\label{sec:method_intro}


\begin{figure*}
\centering
\ForceWidth{3in}
\BoxedEPSF{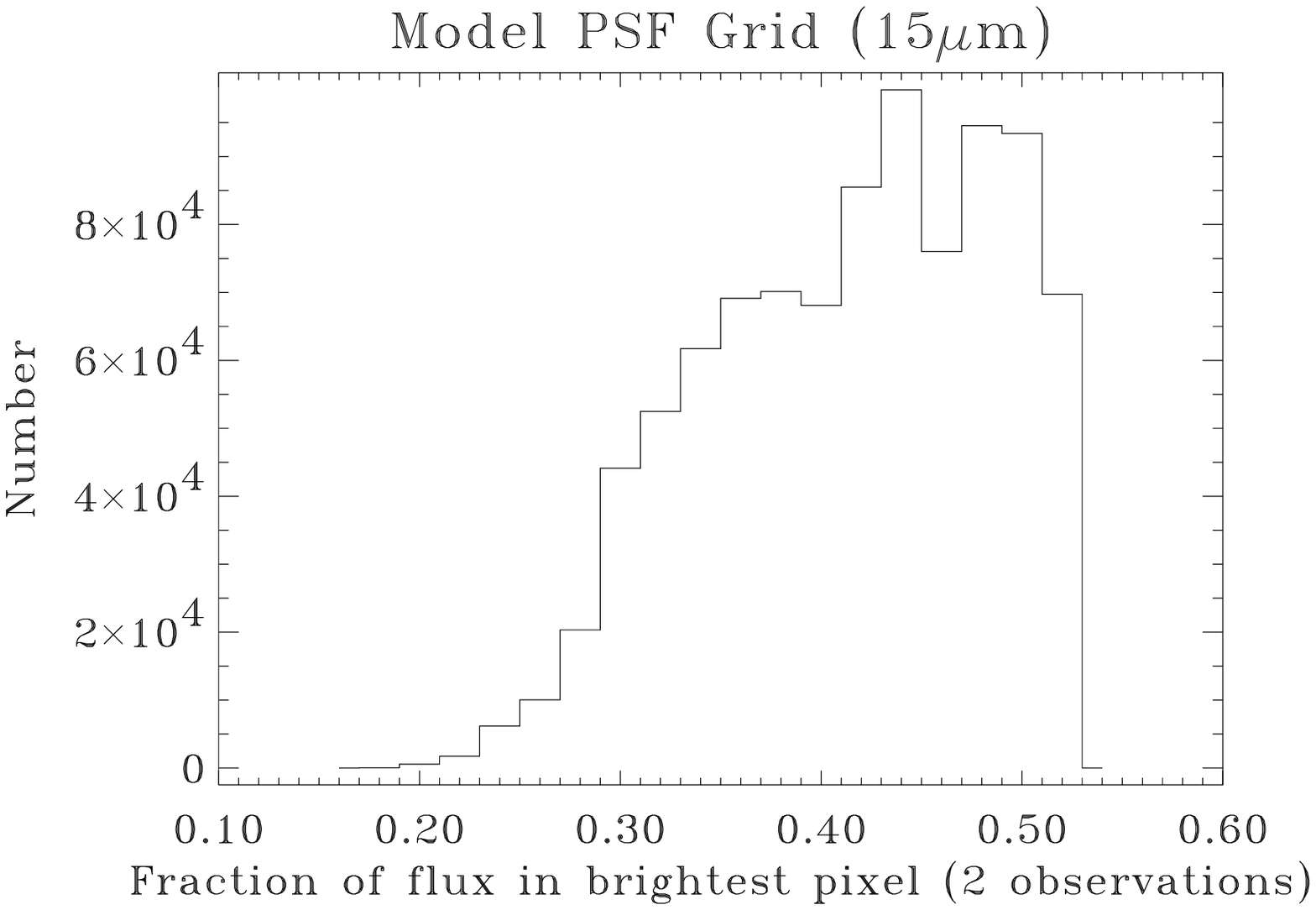}
\ForceWidth{3in}
\BoxedEPSF{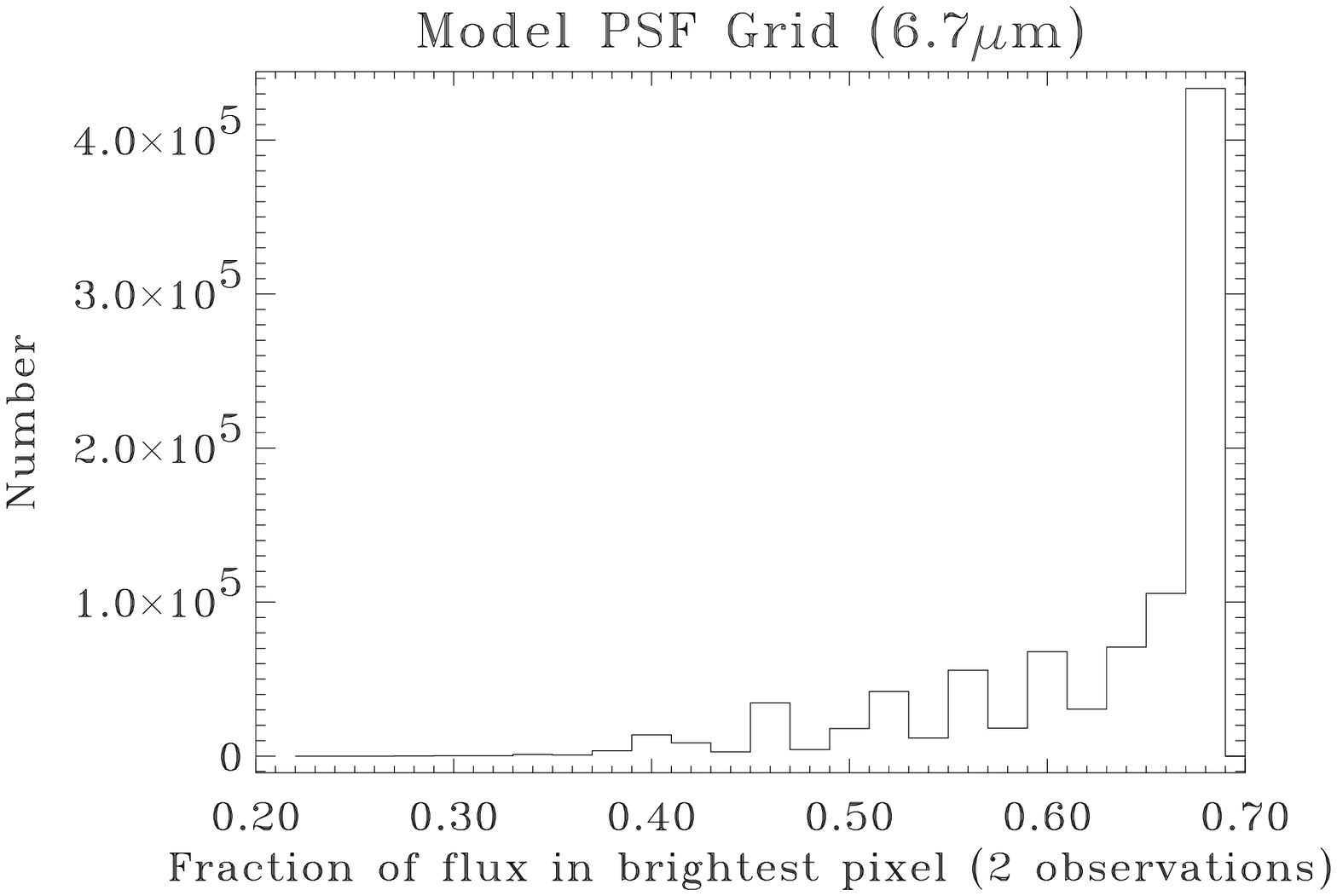}

\ForceWidth{5in}
\vspace*{-1cm}
\hSlide{-3.5cm}
\BoxedEPSF{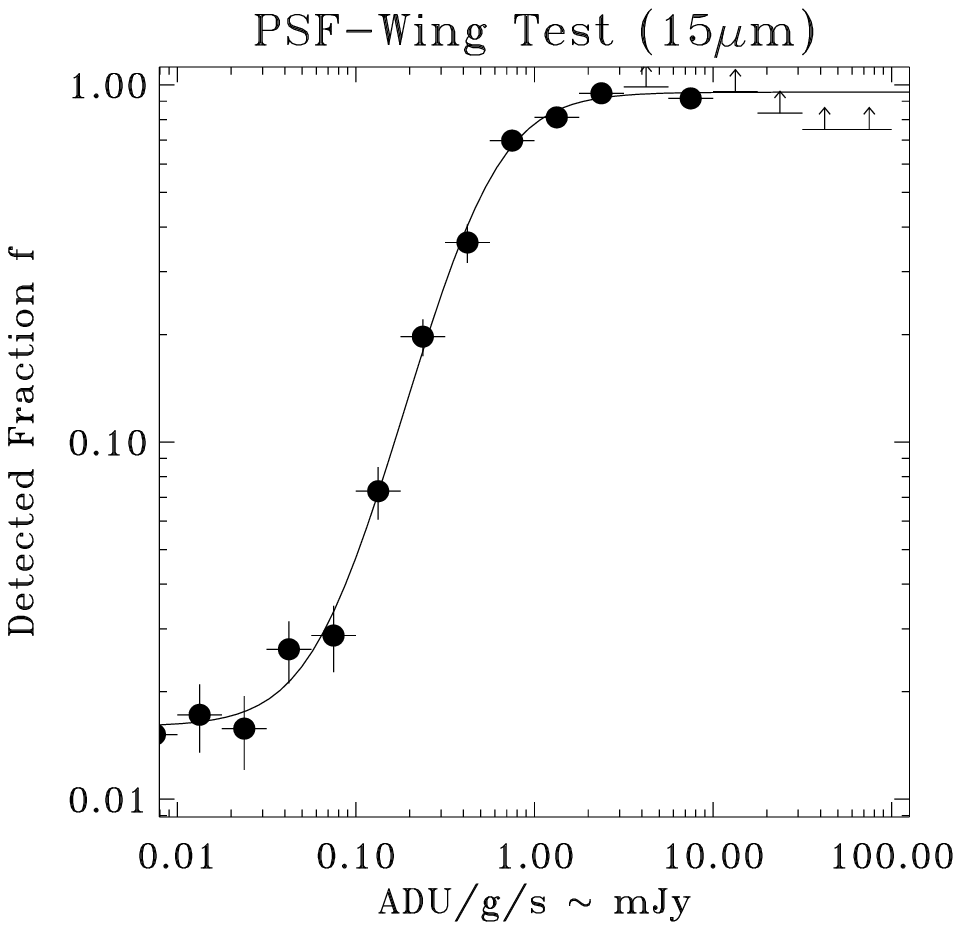}
\ForceWidth{5in}
\vspace*{-9.1cm}
\hSlide{4.5cm}
\BoxedEPSF{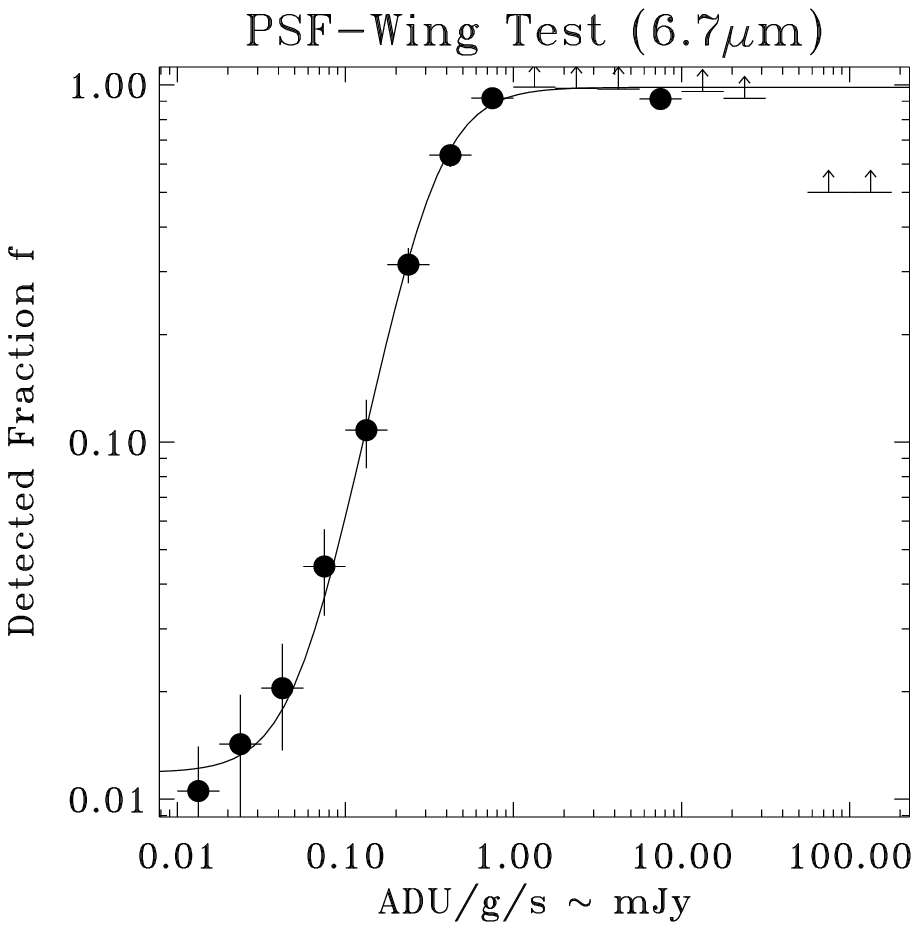}

\ForceWidth{5in}
\vspace*{-2cm}
\hSlide{-3.5cm}
\BoxedEPSF{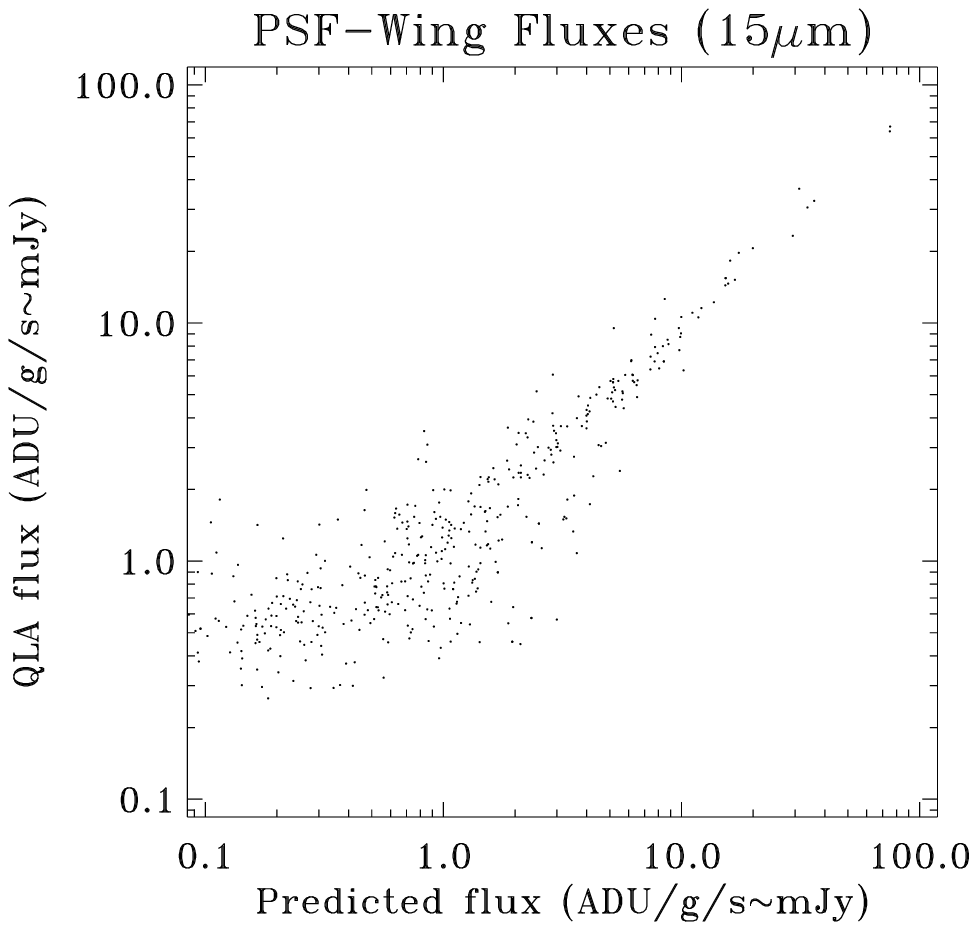}
\ForceWidth{5in}
\vspace*{-9.1cm}
\hSlide{4.5cm}
\BoxedEPSF{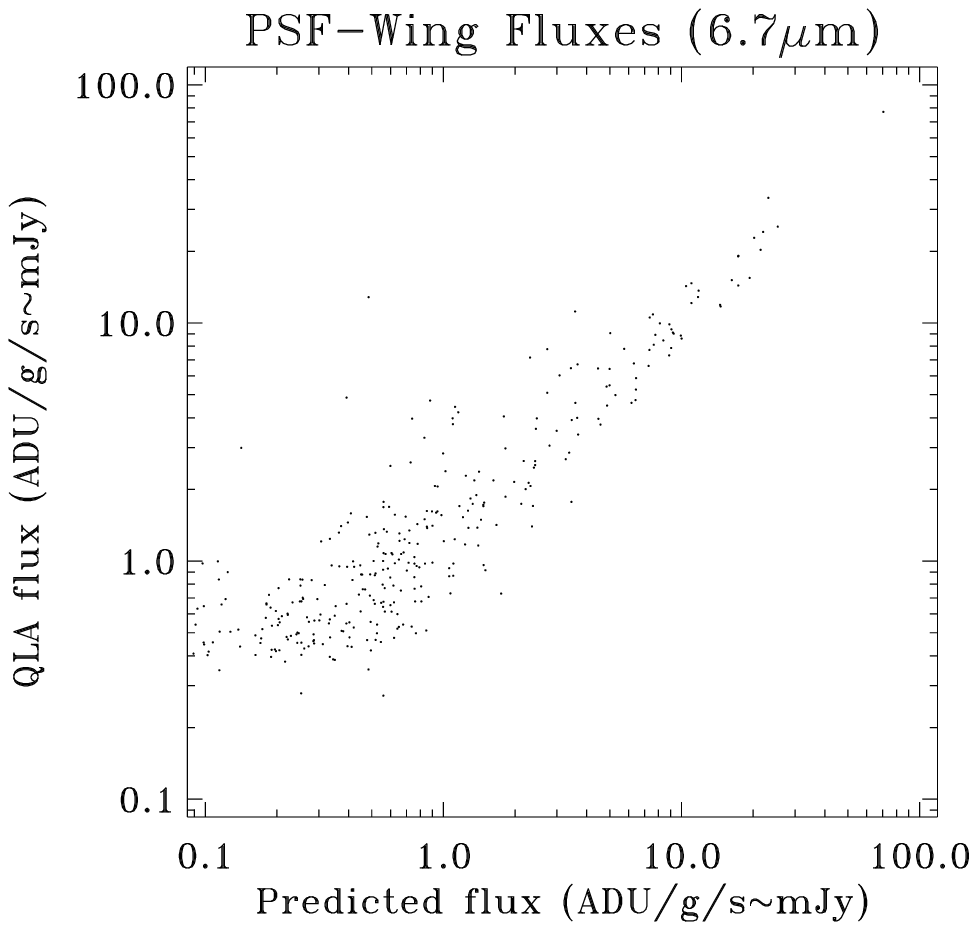}

\vspace*{-1cm}
\caption{\label{fig:fluxcal}
Panels show the $15\mu$m (left panels) and $6.7\mu$m (right panels)
flux calibration data. Top panels show the distribution of the
fraction of flux in the brightest ($6''$) 
pixel, 
out of {\it two} randomly placed observations of a point source. 
To calculate this, we use an 
evenly-spaced grid of theoretical PSF models from Aussel
(priv. comm.).
The distribution is skewed to higher fluxes compared to
the expected peak flux distribution
of a single observation, and has smaller variance.
The central panels show the detected fraction of individual pixels in
the PSF wings of bright 
stellar sources. The central pixels in each star have been
excluded from these plots. 
The lines show the best fit parametric model
using equation \ref{eqn:psf_wing}.
The lower panels show 
comparisons of the detected fluxes in individual pixels in bright
point source PSF wings, with the predictions based on the PSF
model. Again, the 
central pixels have been excluded from these plots.
} 
\end{figure*}

The ELAIS CAM survey proper was conducted in raster mode 
(astronomical template CAM01), with the LW-2 ($6.7\mu$m) and LW-3
($15\mu$m) filters. Details of the CAM01 Astronomical 
Observation Template (AOT) can be found in paper I. 
The CAM detector is stepped across the
sky in a grid pattern, with roughly half-detector-width
steps in one direction and roughly whole detector widths in the
other, covering approximately half a square degree per raster. 
This pattern leads to a redundancy of at least $\times2$ over most of
the area surveyed. 
At each raster pointing (i.e. each grid position of the raster) the
$32\times32$ CAM detector 
is read out several times. 

Like the ISO-HDF North data (Serjeant et al. 1997, 1999, Aussel et
al. 1999,  D\'{e}sert et al. 1999)
the ELAIS CAM data contains many problematic artefacts. 
Because of the frequent and complicated glitches, 
we do
not take the approach of reconstructing a sky-map and searching for
sources in these maps. Rather, we look for the characteristic signatures
of sources and glitches in the time histories of individual pixels.
See Starck et al. (1998) or Aussel et al. (1999) for more details. 

The CAM
detector also 
exhibits hysteresis.
Source fluxes are initially around a
factor of $2$ fainter in instrumental units than the 
stabilised (i.e. asymptotic) value. 
Our survey strategy ensures that sources almost always
have corroborating 
sightings in separate pixels. This permits a filter to
remove glitch events from candidate sources. 

\subsection{Preliminary source extraction pipeline}\label{sec:cam_pipe}
\subsubsection{Preprocessing and deglitching}
The available CAM data reduction software underwent several
substantial improvements over the lifetime of ISO, as the knowledge of
the detector characteristics improved. However, from the outset we
needed a method of preliminary data analysis, to feed for
example immediate
follow-up projects. Such a Preliminary Analysis 
pipeline may not of course
represent best practice at the end of the ISO mission, but should at
least provide reasonably complete and reliable preliminary source
list. It was decided that the CAM Preliminary Analysis 
data reduction should be
as uniform as possible, which required that the 
Preliminary Analysis 
pipeline be fixed at an early stage. Accordingly our
adopted pipeline could not incorporate the accurate field distortion
corrections (Abergel et al. 1998), which were not established at the
start of the mission, 
nor the analytic models for the cosmic ray transients and detector
hysteresis which were developed in the course of the ISO lifetime
(e.g. Lari 1999, Abergel et al. 1998). 
Nevertheless, such improvements will be incorporated in future
ELAIS Final Analysis products.

\begin{figure}
\centering
  \ForceWidth{5in}
\vspace*{-1cm}
  \hSlide{-2cm}
  \BoxedEPSF{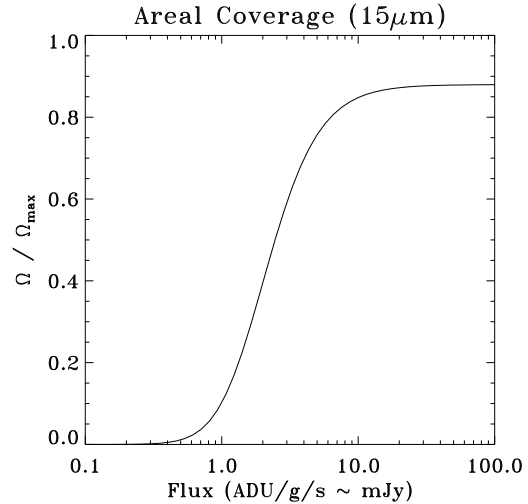}
\caption{\label{fig:areal_coverage_lw3}
Effective areal coverage in the LW-3 $15\mu$m band. The maximum
coverage, which excludes the first and last pointings of each raster and
regions without redundancy (removing $7\%$ from each raster), is
$\Omega_{\rm max}=10.0$ square degrees.  
The curves asymptote to $<1$ because the bright end of the PSF wing
test is not well constrained, having only a few weak limits. 
Also the asymptotic value at low fluxes is not included (and is also
ill-constrained) since this is presumably glitch confirmation. 
} 
\end{figure}

\begin{figure}
\centering
  \ForceWidth{5in}
\vspace*{-1cm}
  \hSlide{-2cm}
  \BoxedEPSF{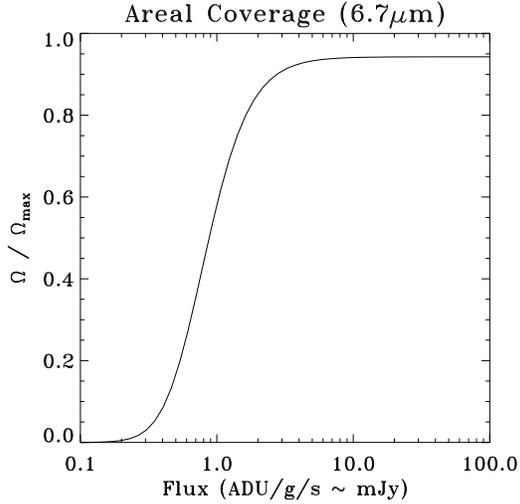}
\caption{\label{fig:areal_coverage_lw2}
Effective areal coverage in the LW-2 $6.7\mu$m band. The maximum
coverage, which excludes the first and last pointings of each raster and
regions without redundancy (removing $7\%$ from each raster), is
$\Omega_{\rm max}=6.51$ square degrees. 
} 
\end{figure}

\begin{figure}
\centering
  \ForceWidth{4.5in}
  \hSlide{-1.5cm}
  \BoxedEPSF{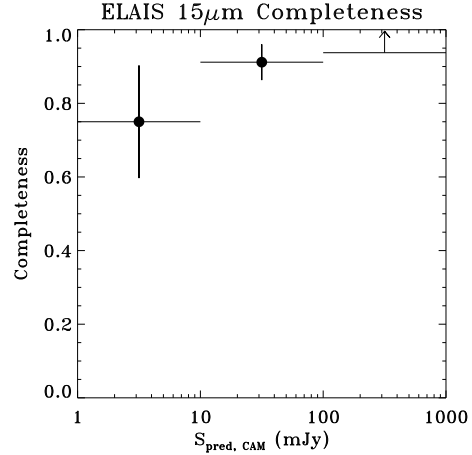}
\caption{\label{fig:comp_rel_hip_completeness}
Completeness at $15\mu$m estimated from the predicted fluxes of stars
in the main ELAIS regions. For more details see Crockett et al. 1999. 
}
\end{figure}

\begin{figure}
\centering
  \ForceWidth{4.5in}
  \hSlide{-1.5cm}
  \BoxedEPSF{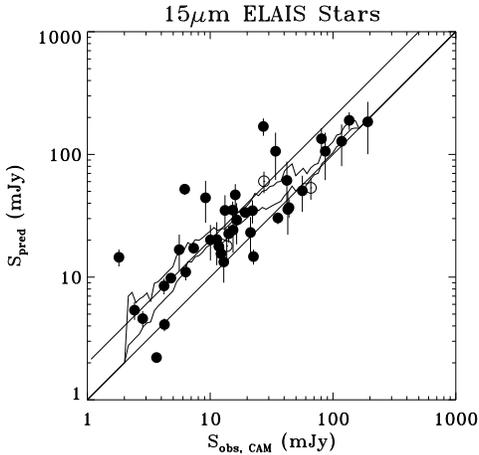}
\caption{\label{fig:comp_rel_hip_fluxcal}
Flux calibration at $15\mu$m estimated from the predicted fluxes of stars
in the main ELAIS regions. The observed fluxes assume a one-to-one
conversion between ADU/gain/second and mJy. Open points are from
Hipparcos, and solid points from Simbad. As discussed in Crockett et
al. 1999, the most reliable spectral typing and hence mid-IR
predictions are available for Simbad stars. The straight lines show a
$1:1$ and $2:1$ ratio between observed and predicted counts. Also
shown is the $\pm1\sigma$ limits on the mean observed:predicted ratio
as a function of flux, calculated within $\pm0.25$ dex of each flux
(i.e., an $0.5$ dex boxcar smoothing). There are perhaps hints from this
figure that the flux calibration may be a weak function of flux, though in
this paper we assume a flux-independent scaling. 
For more
details see Crockett et al. 1999 and Misoulis et al. 1998. 
}
\end{figure}

The data reduction was performed using the Interactive Data Language
(IDL) software. The initial steps of the CAM Preliminary Analysis pipeline are
straightforward. The edited 
raw data supplied by ESA were converted to IDL structures 
using the CAM Interactive Analysis (CIA, April 1996 version), and
converted from ADUs (analogue-to-digital units) to ADUs per second. 
The default dark frame was subtracted from each exposure. To estimated 
the noise level in each pixel, we performed an iterative Gaussian fit
to the histogram of readout values for each pixel.
Cosmic ray spikes were then identified by 
$>4\sigma$ rises followed (one or two readouts later) by $>4\sigma$
falls. A similar algorithm was used to identify occasional readout
troughs. The readout histograms were then re-fit and an empirical sky
flat field was obtained; the default ESA-supplied flat field was found
to give very unsatisfactory results. An attempt was made to model the
initial detector stabilisation using the IAS model within the CAM
Interactive Analysis (CIA) software package. 

\subsubsection{Background estimation and Source Detection}

Unlike the detectors on the IRAS satellite (e.g. Neugebauer et
al. 1984), the transients in the 
ISOCAM detector make the background levels in each pixel vary strongly
and discontinuously with time.
The 
approach adopted in the Preliminary Analysis was to  
estimate the background level in a given pixel and pointing from
linear fits (in time) to the readouts in previous and subsequent
pointings, then identify candidate sources from $>3\sigma$ features
above the background in the pixel readout timelines.  
The source extraction pipeline is therefore spatially $1$ dimensional.
This simple approach avoids any
explicit parameterisations of the transient profiles, which were not
available at the time, and allows a
local error estimate of the background level. This error was taken to
be the formal error on the fit, but did not incorporate the
instrumental noise at each data point. 
Since cosmic
rays were frequently observed to cause discontinuities, the range for
the linear fits extended no further than the nearest (in time) cosmic
ray, and in any case not longer than $3$ pointings. If an acceptable
$\chi^2$ was not obtained in the linear fit, the range for the fit was
decreased; if a good fit was still not obtainable, an average of the
$10$ nearest readouts was taken. Sources could then be identified from
their excess above the extrapolated backgrounds. Since sources also
create discontinuities, the data stream was iteratively re-fit treating
$>3.5\sigma$ sources in the same way as cosmic rays above. An initial
list of brightest sources was obtained before the iterative fitting,
by searching for their discontinuous 
rise at the start of a pointing, and discontinuous fall at the end. 
Where sources were found (whether initially or in the iterative
fitting) the background level in later iterations was extrapolated
from only the previous pointings, ignoring the subsequent pointings,
since the hysteresis after a source would otherwise lead to an
overestimation of the 
background level. Sources were not extracted from
the first or final pointings of the rasters, owing to limitations in
the background fitting routines at the time the pipeline was frozen. 
Note that this iterative source extraction does not distinguish
genuine sources from ``fader'' transient events, so a further 
filter for source candidates is still required. 

\subsubsection{Source Corroboration}
The source detections in the pixel readout histories were spatially
merged in 
each pointing using a connected pixel algorithm. (Note that there was
no minimum number of pixel detections, unlike e.g. connected pixel
source detection on an oversampled CCD image.)
We then used the
$\simeq\times2$ redundancy in the CAM rasters to search for
corroborating observations of each source candidate. Genuine sources
should be present in both observations, but glitch events should not
be confirmed except by chance. 
The adopted 
search radius of $2$ pixels, while large enough to safely encompass the
(then uncertain) field distortion, nevertheless led to a large
number of spurious detections, with the majority of source candidates
at $15\mu$m due to glitch corroboration. 
Each candidate corroborated source was
therefore examined by eye independently by at least two observers, who
assigned quality flags of $1-4$, where $1$ refers to a ``definite
source,'' $2$ is a ``probable source,'' $3$ ``probably not'' and $4$
``definitely not.'' 
Each raster in the LW-2 ($6.7\mu$m) filter yielded
typically $\sim100$ events in total of which around one half were
classified as probable sources by at least one observer. 
In LW-3 ($15\mu$m)
both the number and fraction of spurious events was much higher:
there were typically $\sim30$ strong source candidates, and a further
$30-50$  sources where the classification was ambiguous or debatable,
with typically around $200-500$ spurious events. 
Note that the Preliminary Analysis  
algorithm will necessarily miss sources with
(a) only one observation or (b) corroborating observations in only the
first or last pointings, so the effective area is slightly less than the
nominal $\sim12$ square degrees. 

\subsubsection{Astrometric corrections}
After the Preliminary Analysis reduction was complete, we improved the astrometry by
incorporating the latest field distortion correction into the $15\mu$m
source catalogues. 
Several sources with strong transient
events nearby had their centroids strongly affected by the glitches.
We therefore adopted a simple strategy for our Preliminary Analysis
astrometry 
and flux calibration: the flux and (distortion-corrected) position of
a source are taken from 
those of the brightest single-pixel detection of that source,
excluding transients. 
We found this to be superior to (eg) masking nearby transients by eye 
then recalculating the centroids of the eyeball-accepted sources, 
particularly if the PSF wings lie on the detector but the source
itself is just outside.  
Our adopted  algorithm 
should yield astrometry accurate to $\pm3''$ in the absence of any
other systematic errors. 
Two such systematics were expected in our data:
firstly, errors in the position of the lens introduce a random
astrometric offset to each raster of order one pixel; secondly, any
errors in the calculation of 
the pointing position by CIA would offset any sources in that
pointing. By examining the offsets with the likelihood-ratio
identifications we can determine the lens offset empirically (section
\ref{sec:optid}). However, several rasters were found to have bimodal
distributions 
of ISO--optical offsets, due to some unknown error in the CIA-derived
astrometry in at least part of the raster. We therefore
rederived the pointing astrometry using the ESA-supplied IIPH.FIT
astrometry file, using the median coordinate positions in the duration
of the pointing. This was found to remove the bimodality.

\subsubsection{Aperture Corrections to Photometry}
As discussed above, our source extraction method involves
looking for characteristic time signatures in individual
pixels. Without (at the time) a reliable and exact model for the
glitch events, nor a reliable glitch event identification, we found
that aperture photometry around our source positions was often
seriously affected by nearby glitches. Instead of aperture photometry, 
we simply took the brightest flux of the pixels detecting the source,
excluding (by eye) those pixels affected by glitches. Clearly, some
aperture correction is needed to correct these peak pixel fluxes to
total fluxes. 

We can quantify these aperture corrections using a PSF model. 
In figure \ref{fig:fluxcal}
we show the predicted flux in the
brightest pixel of two randomly positioned observations of a point
source (recall that at least two observations of a source are
required for it to pass the Preliminary Analysis selection). At $15\mu$m, the brightest
pixel has a flux of $\sim0.4\pm0.1$ 
times the total flux of that source. At $6.7\mu$m the histogram is
more sharply peaked, since the PSF is undersampled; the peak flux is
always less than $0.69$ times the total at $6.7\mu$m, but is greater
than $0.5$ ($0.6$) of the total in $>90\%$ ($>75\%$) of occasions. 
We therefore applied global aperture corrections of $2.36$ at $15\mu$m
and $1.54$ at $6.7\mu$m to our peak fluxes. 

\section{Results}\label{sec:results}

\subsection{Eyeballing results}

Our eyeballing results imply our catalogue is highly reliable to at
least $3$ mJy at $15\mu$m. 
In the ELAIS areas considered in this paper, there were $715$ $15\mu$m
sources accepted by two observers, of which $510$ had APM
identifications (section \ref{sec:optid}); of the $816$
singly-accepted sources at $15\mu$m, $212$ had APM
identifications. The singly-accepted sources are heavily skewed to
faint fluxes: $90\%$ are fainter than $2.2$ mJy. 
If we choose to regard all sources accepted by only one observer as
false positives, and all sources accepted by two as true positives,
then we obtain $\sim80\%$ reliability for sources accepted by any
observer at $\sim3$ mJy and 
$\sim95\%$ at $5$ mJy. These are one of the most pessimistic
assumptions we 
could make; if alternatively we merge the doubly-accepted sources with 
the optically identified singly-accepted sources, 
the reliability
(fraction doubly-accepted) of
the sources accepted by any observer 
rises to $\sim 95\%$ 
at $\sim 3$ mJy and increases at brighter fluxes. 
The blank-field singly-accepted sources are heavily skewed to faint
fluxes, so we can neglect their effect on the counts above $3$
mJy and combine the doubly-accepted sources with the optically
identified singly-accepted for the purposes of the source counts. 
However this is not to say that the fainter singly-accepted
sources necessarily have lower reliability, since the 
genuine fainter sources may have fainter optical counterparts. 
The situation is similar at $6.7\mu$m. Of the singly-accepted sources
$95\%$ have fluxes less than $1.5$ mJy. 

\subsection{Completeness, reliability and flux
calibration}\label{sec:comp_rel} 

\subsubsection{Repeated observations}\label{sec:repeat}
There are several potential approaches to estimating the Preliminary Analysis
completeness and reliability, the most obvious of which is repeated
observations over small areas. Oliver et al. (1999) present $10$
repeat observations of a small ELAIS raster, covering at least six
known sources by design. The source extractions in the individual
reductions confirmed our
result that the source extraction is highly complete and reliable at
flux densities above $\sim2.5$mJy at $15\mu$m. 

\subsubsection{PSF-wing test}\label{sec:psf_test}

One robust test of our temporal source extraction is to examine the
point spread function wings of bright sources. 
Like the cross-correlation with bright stars, this relies on the
knowledge that a given pixel is illuminated by a known source flux,
but has the advantage that the comparison can be taken to arbitrarily 
faint flux levels. We selected the $30$ brightest $15\mu$m sources
with stellar 
optical identifications, and 
selected a model theoretical Point Spread Function (PSF; Aussel,
priv. comm.) with the smallest RMS difference in the central
$3\times3$ PSF pixels. 
We normalised the PSF to the source flux using the mean observed
flux in the central pixel. Using this model we predicted the mean flux
in each pixel of the PSF wings, and hence determined the single-pixel
detection 
efficiency $f$ of the temporal source extraction. The results are
shown in figure \ref{fig:fluxcal}.
We can also compare
the predicted PSF fluxes with the fluxes extracted by the Preliminary Analysis
algorithm, shown in figure \ref{fig:fluxcal}. 
The relation is
encouragingly linear, though the scatter is larger than expected 
(by $\sim\times2$ at $15\mu$m) 
based on the quoted errors in the Preliminary Analysis sky background
fits. This is perhaps not surprising, since the background fitting
algorithm does not make use of the detector noise in the fit,
so will tend to underestimate the background level uncertainty; also,
non-white noise features may prevent the noise scaling as
$\sqrt{N}$. 
Oddly, the discrepancy in the scatter is largest at brightest
fluxes. Plausible explanations include slight errors in the source
centroids or in the theoretical PSF shape, both of which sensitively
affect the brightest flux predictions. 
Analogous results for the $30$ brightest $6.7\mu$m stars are
also shown in figure \ref{fig:fluxcal}. 
The undersampling of the PSF at $6.7\mu$m makes the scatter 
harder to interpret: more of the
flux is contained in the central pixel 
making it less sensitive to errors in the
assumed PSF shape, but is more affected by the much larger uncertainty
in the centroid. 

There are several caveats which apply to the high apparent
completeness in figure \ref{fig:fluxcal}. 
The chance
detections of ``faders'' will tend to 
increase this estimate, but 
since both the $6.7\mu$m and $15\mu$m completeness appear to asymptote
to $\sim0.015$ (i.e. probability of detection of nearby spurious events
is $1.5\%$) this appears not  
to be a serious problem. This is also only 
applicable to single-pixel detections, whereas in fact the CAM
detector Nyquist samples the PSF at $15\mu$m. Indeed at $15\mu$m the
fraction of the flux in 
the brightest pixel rarely exceeds $0.5$ 
though the same does not apply at $6.7\mu$m. 
The completeness also is not 
the detected source fraction $F$ but rather $F^2$, since we require at
least $2$ 
independent detections for a source to be accepted by our algorithm. 
There is 
also a slight bias in that the brightest detected sources are
not typically observed when the detector is suffering the strongest
transients, because the sources would not otherwise be
detected. Finally, any incompleteness caused 
by the eyeballing, or by any areal coverage lost to e.g. cosmic ray
impacts, would not be included in these estimates as they stand. 

We can therefore estimate the Preliminary Analysis completeness from the PSF-wing test in 
the following way. We 
can convolve the single-pixel detection rate in
figure \ref{fig:fluxcal}
with the peak flux distribution
expected from the PSF models 
to obtain
the source sensitivity $F$ for single observations. The Preliminary Analysis
completeness, before eyeballing, will then be proportional to 
$F^2$, assuming the PSF wings themselves are 
representative of the data as a whole. 

We fit the detected fractions in figure
\ref{fig:fluxcal}
with $\tanh$ functions, i.e.
\begin{equation}\label{eqn:psf_wing}
f(S) = \frac{(f_{\rm max}-f_{\rm min})}{2}\tanh(\alpha\log_{10}(S/S_0)
+ 1) + f_{\rm min} 
\end{equation}
where $f_{\rm min}$ and $f_{\rm max}$ represent the asymptotic limits
at faint and bright fluxes respectively.
We define the single-pixel source detection fraction to be
$f'(S)=f(S)-f_{\rm min}$, and use a grid of PSF models (each
normalised to 1) spanning the possible range of centroid positions
to estimate the single-pointing source detection:
\begin{equation}
F(S) = \frac{1}{N}\sum_{i=1}^{N} f'(S\times S_{\rm peak, i})
\end{equation}
where $S_{\rm peak, i}$ is the peak flux in the $i$th PSF. 
This assumes that if a source is not detected in the peak pixel, it
will not be detected in any pixel. 
The maximum areal coverage of the Preliminary Analysis, ie excluding the first
and last pointings of each raster, and regions with no redundancy, is
$\Omega_{\rm max}=10.0$ square degrees at $15\mu$m, and $\Omega_{\rm
max}=6.51$ square degrees at $6.7\mu$m. 
From this we obtain the Preliminary Analysis areal coverage as a function of flux:
\begin{equation}
\Omega_{\rm Preliminary Analysis}(S) = \Omega_{\rm max} F^2(S)
\end{equation}
The final areal coverage from the PSF-wing test is plotted in
figures \ref{fig:areal_coverage_lw3} and 
\ref{fig:areal_coverage_lw2}.

\subsubsection{Comparison with IAS and CEA pipelines}\label{sec:amanda}
As a final check of the completeness of our Preliminary Analysis catalogue, we compared
our source extraction in the repeat observation regions (c.f. Oliver
et al. 1999 and section \ref{sec:repeat}) with extractions
made with the CAM-Deep pipeline developed at the Commisariat \'{a}
L'Energie Atomique, Saclay (CEA; Baker,
priv. comm.) and a pipeline based on the ``Triple Beamswitch'' method 
developed at Institut d'Astrophysique Spatiale (IAS; Clements,
priv. comm; D\'{e}sert et al. 1998). Of our six robust sources
(section \ref{sec:repeat}), CEA 
and IAS both identify 
three (the same three), with fluxes brighter than $\sim3$mJy. This
appears to be mainly because the Preliminary Analysis pipeline 
extracts lower signal-to-noise sources, but supplements with greater
manual eyeballing. Nevertheless, this confirms that we are not
significantly underestimating the surface density of sources brighter
than $3$mJy.

\subsubsection{Flux calibration and Cross-correlation with bright
stars}\label{sec:hipparcos} 
Another useful test of the completeness is to search for detections at
the locations of bright stars (see section \ref{sec:optid} for details
of the optical identification algorithm). If
the spectral types of the stars are 
known, one can predict their mid-infrared fluxes. 
An exception would be dust-shell stars, but these are expected to be
rare in the survey. In Crockett et al (1999) and Crockett (1999) the
sources are 
cross-correlated with stars from the Simbad and Hipparcos
databases. All the $22$ stars with predicted fluxes brighter than
$3$mJy at $15\mu$m were detected in the Preliminary Analysis, and all but two of the
$20$ stars with predicted $6.7\mu$m fluxes above $1$mJy appeared in
the Preliminary Analysis. To assess the level of random associations, we randomised
the stellar positions within the ELAIS fields and repeated the
cross-association, and found none. The $15\mu$m completeness is shown
in figure \ref{fig:comp_rel_hip_completeness}. Note that this is an
extremely robust estimator of the completeness, since the stellar flux 
prediction algorithm reproduces IRAS fluxes well, and the stars
cannot be argued to lie on atypical regions of the detector. 

However, the flux calibration implied by these associations is around
a factor of $2$ discrepant with the expectation that ADU/gain/second
$\simeq$ mJy at $15\mu$m (see figure
\ref{fig:comp_rel_hip_fluxcal}). Across the entire range in flux, the
$\pm1\sigma$ limits on the log of the calibration ratio are 
$0.246\pm0.050$, ie the ratio is $1.75^{+0.26}_{-0.23}$. As shown in
figure \ref{fig:comp_rel_hip_fluxcal}, there are hints that this
calibration is a function of flux, with a calibration ratio of $2$
preferred at faint fluxes (where indeed most of our sources lie). 
(The ISOCAM Observer's Manual
recommends a conversion of approximately $2$ ADU/gain/second to $1$
mJy at $15\mu$m for fully stabilised sources, and after correcting for 
the loss of flux due to lack of stabilisation (e.g. 
section \ref{sec:method_intro}) becomes
around a $1:1$ conversion.) To predict the fluxes 
we followed the procedure of Crockett et al. (1999) and Misoulis et
al. (1998), incorporating the passband profiles (for more details we
refer the reader to these papers; the algorithm accurately reproduces
stellar mid-IR IRAS fluxes so there are unlikely to be significant
systematics in the CAM flux predictions). It is not immediately clear
what might cause such a discrepancy, though there are several
possibilities, such as the uncertainties in the PSF for the aperture
correction, and the assumed $\sim\times2$ loss in flux (section
\ref{sec:method_intro})  from the lack
of an upward source stabilisation correction. 
(Note that differences in the PSF due to the differing spectral slopes 
of the stars are too small to account for the discrepancy: e.g. Aussel
et al. (1999) find it only to be a $\sim10\%$ effect.)
For the purposes of the
source counts we will adopt the mJy : ADU/g/s = $1:2$ stellar
calibration implied in figure \ref{fig:comp_rel_hip_fluxcal}, where
the ADUs are not corrected for losses due to lack of stabilisation.

At $6.7\mu$m a lower
bound of $95\%$ can be made on the completeness at fluxes $>10$ mJy,
but the uncertain aperture corrections make applying the stellar flux
calibration more difficult. As a preliminary measure we therefore
simply take the pre-flight 
ISOCAM Observer's Manual calibration at $6.7\mu$m, corrected by a 
factor $\sim2$ (section \ref{sec:method_intro}) to account for the
loss in flux from lack of 
stabilisation.  

In summary, our various completeness estimates yield a $\geq 1$ mJy
limit to the completeness at $15\mu$m, and $\geq 0.5$mJy at $6.7\mu$m
(figures \ref{fig:areal_coverage_lw3} and
\ref{fig:areal_coverage_lw2}) using the ISOCAM observer's manual flux
calibration corrected by a factor of $2$ to account for stabilisation
loss (section \ref{sec:method_intro}). However our stellar
cross-correlation suggests we have underestimated the fluxes by a
factor of $\sim2$ at $15\mu$m (figure \ref{fig:comp_rel_hip_fluxcal})
so the $15\mu$m completeness quoted should by revised upwards to 
$\geq 2$ mJy. In {\it all} subsequent discussion, the $15\mu$m ELAIS
fluxes are assumed to obey this stellar flux calibration. 


\begin{figure*}
\vspace*{-3cm}
\centering
  \ForceWidth{3in}
  \hSlide{-1cm}
  \BoxedEPSF{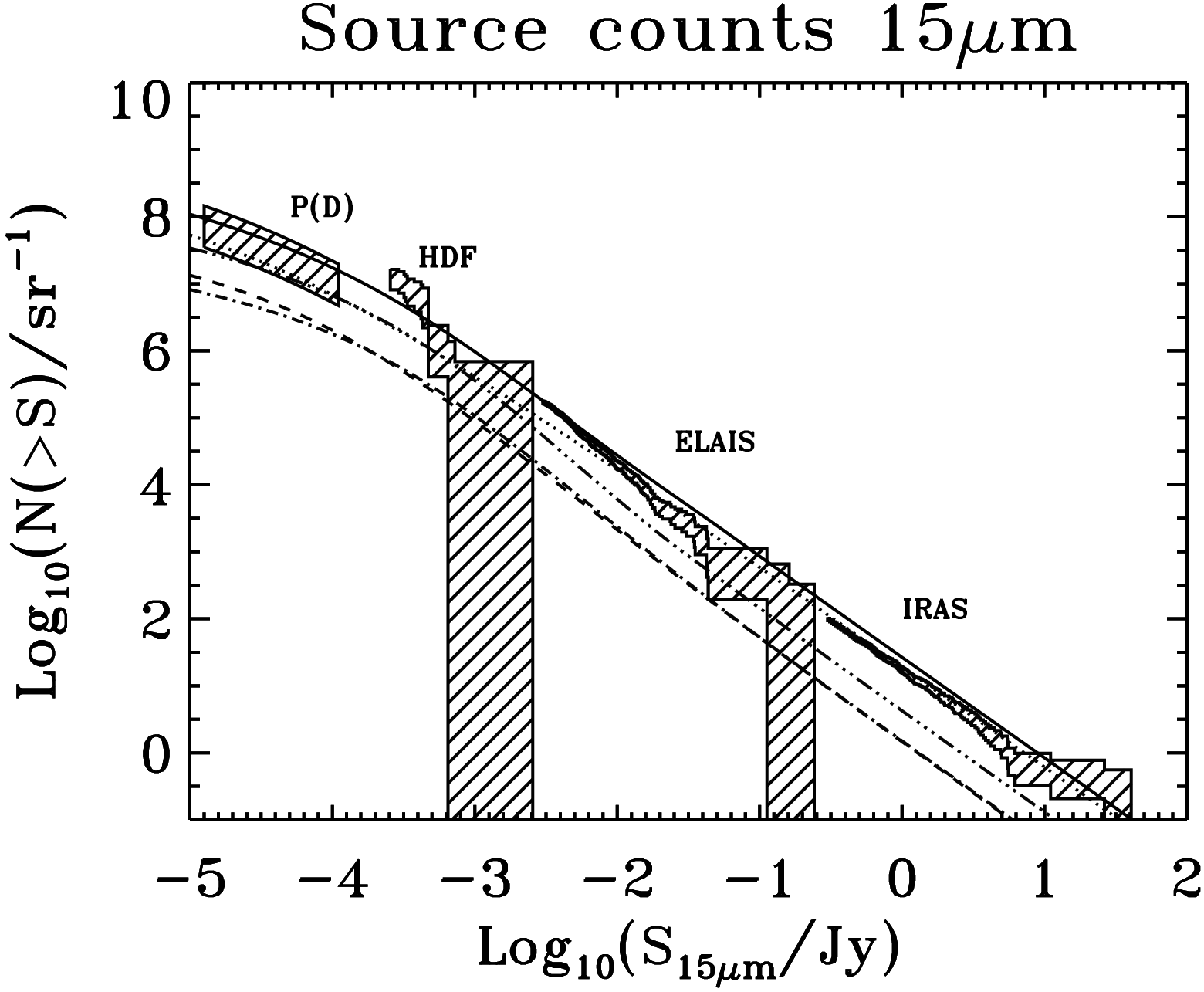}
  \ForceWidth{3in}
  \hSlide{0cm}
  \BoxedEPSF{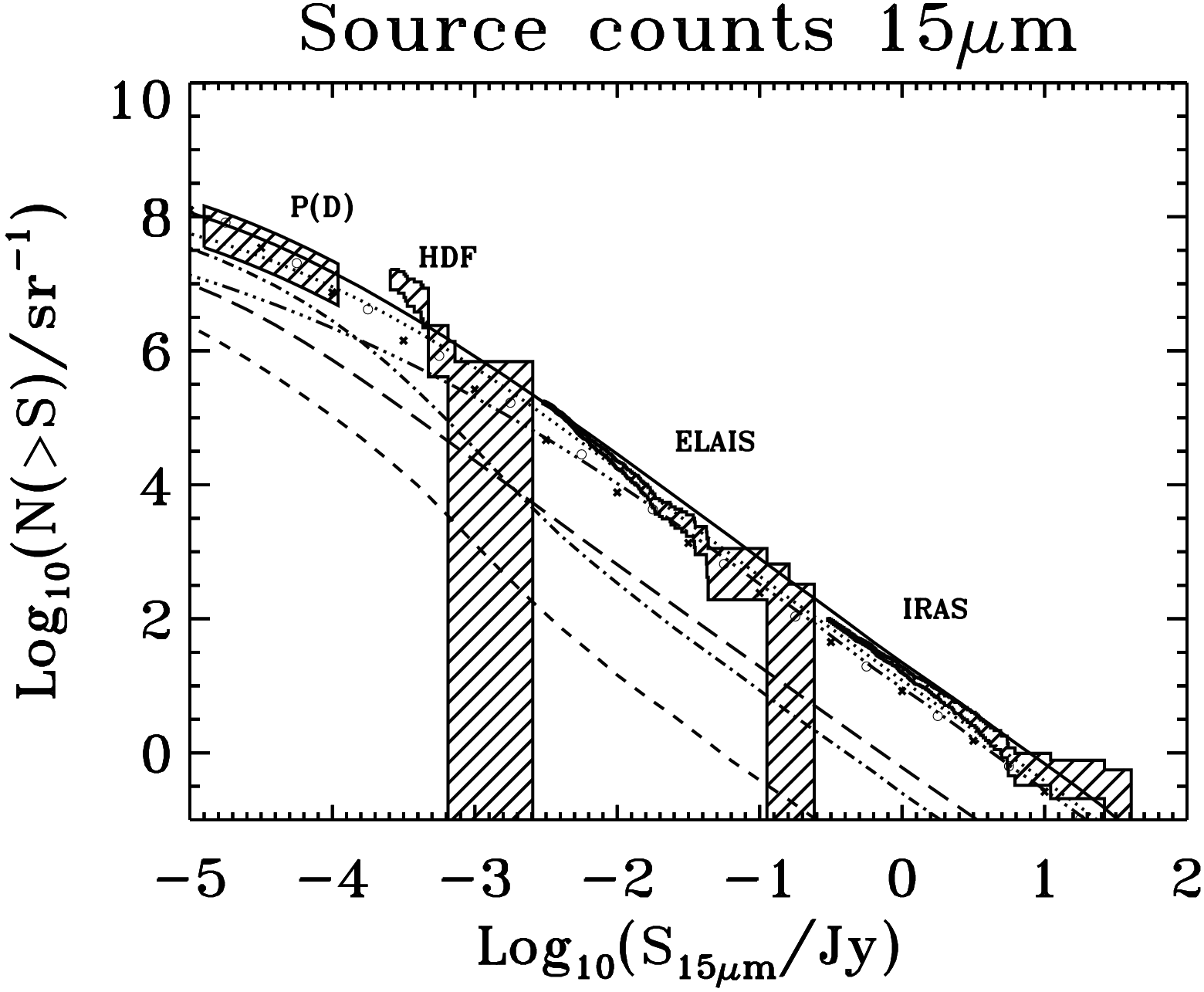}

  \vspace*{-2cm}
  \ForceWidth{3in}
  \hSlide{-1cm}
  \BoxedEPSF{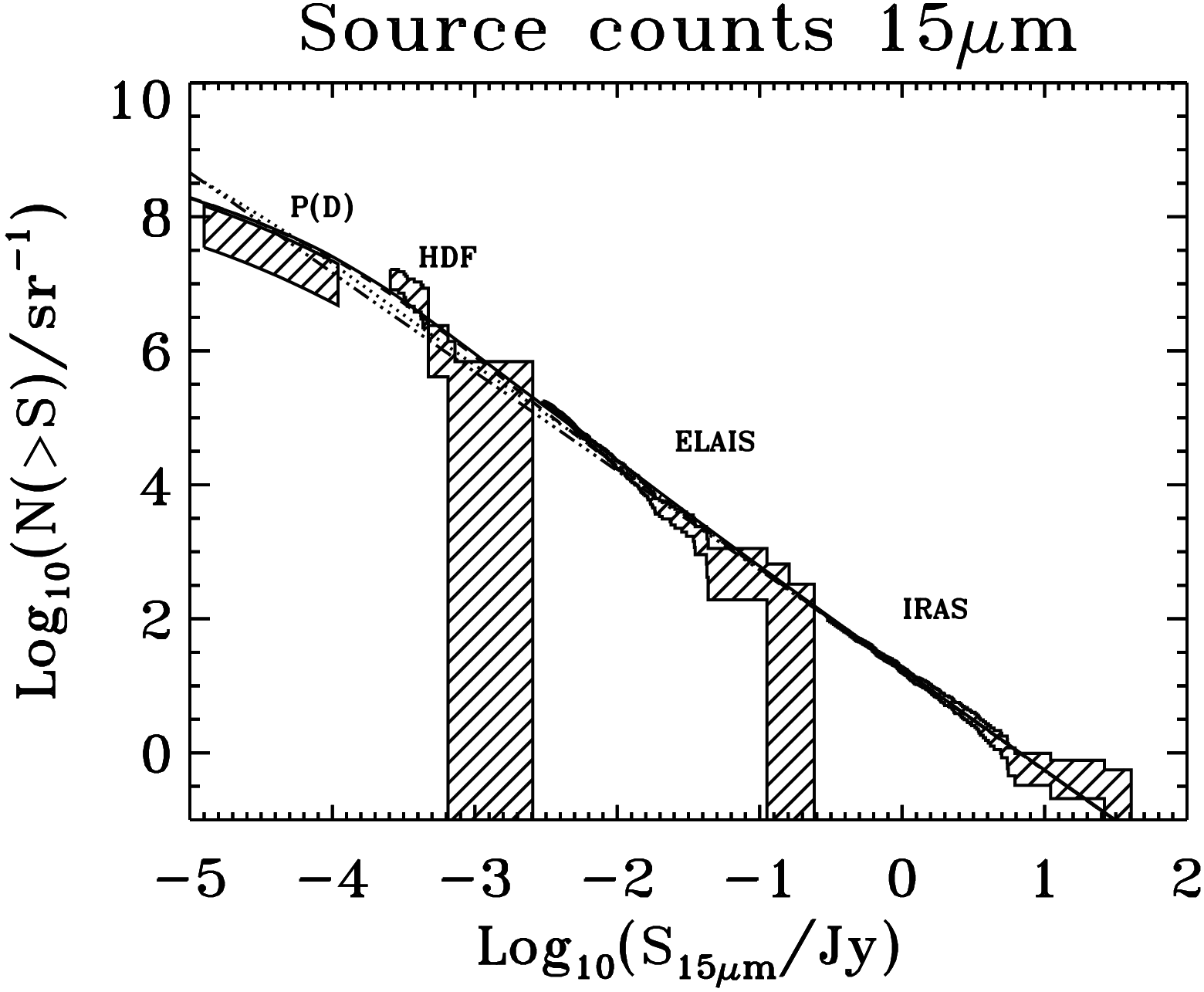}
  \ForceWidth{3in}
  \hSlide{0cm}
  \BoxedEPSF{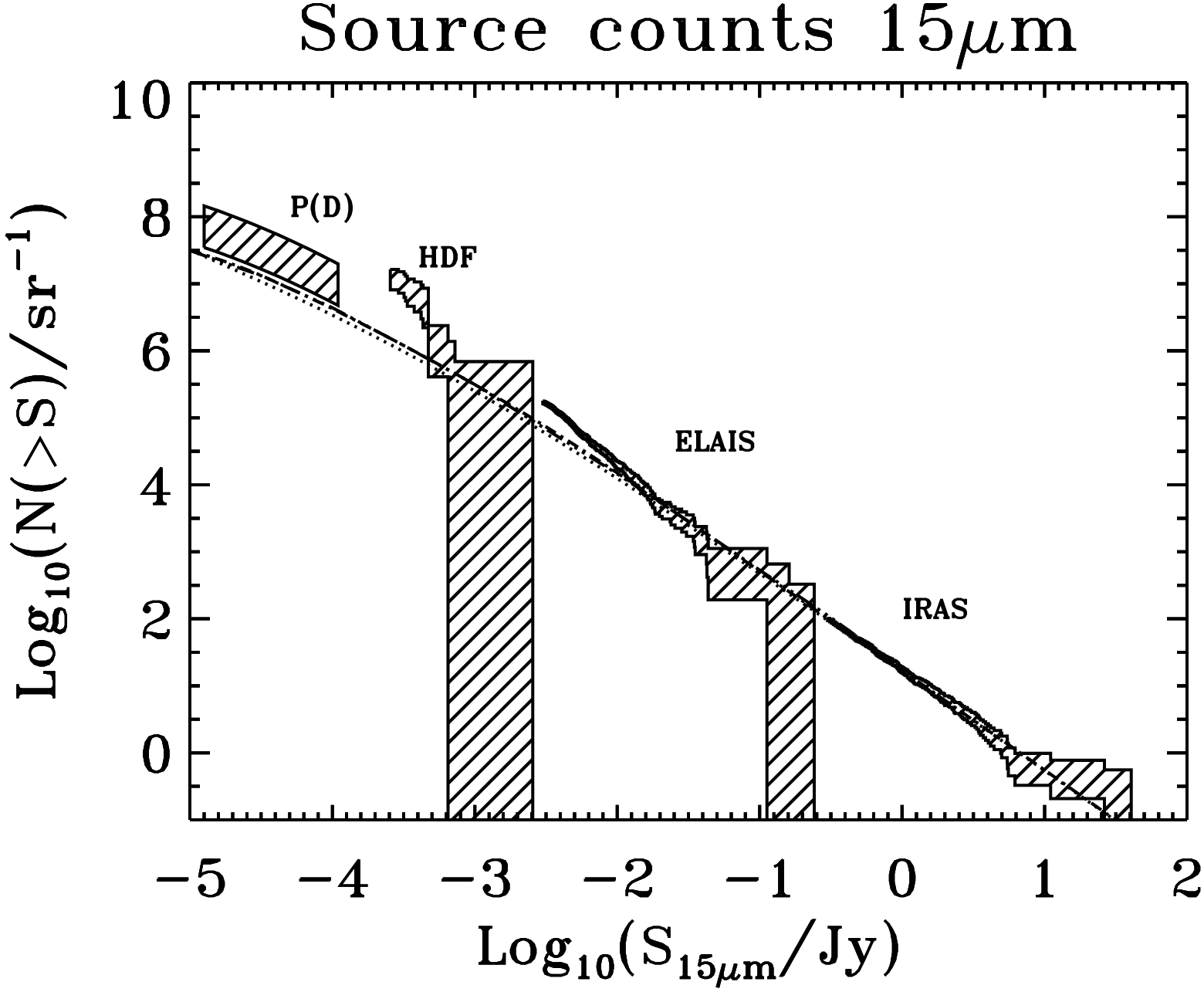}
  \vspace*{0.5cm}
\caption{\label{fig:counts}
Integral ELAIS extragalactic source counts at $15\mu$m. Flux densities
are quoted in Jy.  
Shaded regions show the ranges spanned by $\pm1\sigma$ uncertainties.
Also shown are the source counts and P(D) analysis from the Hubble
Deep Field (Oliver et al. 1997, Aussel et al. 1999).
The IRAS counts are estimated from
the $12\mu$m counts as discussed in the text. 
Top Left panel shows the 
Pearson \& 
Rowan-Robinson model. 
Dotted line shows normal galaxies,
dashed line shows Seyfert I galaxies, dot-dashed are Seyfert IIs, and
starbursts are dash-dot-dot-dot. The total is shown as a full line. 
Top Right panel shows the Franceschini et al. (1994) source count model
overplotted.  
The model spiral contribution is shown as a dotted line, ellipticals
as a dashed line, S0 as a dash-dot line, starbursts as a
dash-dot-dot-dot line and AGN as a long dashed line. 
The total population model is shown as a full line. 
Also plotted are the Guiderdoni et al. (1997) models A and E, as
small filled crosses and small open circles respectively. 
Bottom Left panel shows the evolving Xu et al. (1998) models,
renormalised by 
$\times1.8$ to match the
IRAS counts as in figure \ref{fig:counts}.  
The full line shows $(1+z)^3$ luminosity evolution with K-corrections
derived from starburst models with mid-IR features, and the dashed
line shows the same luminosity evolution without the mid-IR
features. Density evolution of $(1+z)^4$ with mid-IR features is
plotted as a dotted line, and without mid-IR features as a
dash-dot-dot-dot line. 
Bottom Right panel shows all available no-evolution models. 
Franceschini et al. (in prep.) is 
overplotted as a dash-dot-dot-dot line. 
The Xu et al. (1998) models are shown with 
and without the MIR spectral features (dash-dot and dotted 
respectively). All three no-evolution models have been renormalised
to match the IRAS counts, by a factor of $0.8$ for the Franceschini et 
al. models, and $1.8$ in the case of the Xu et al. (1998) models. 
}
\end{figure*}

\begin{figure*}
\centering

  \vspace*{-2cm}
  \ForceWidth{3in}
  \hSlide{-1cm}
  \BoxedEPSF{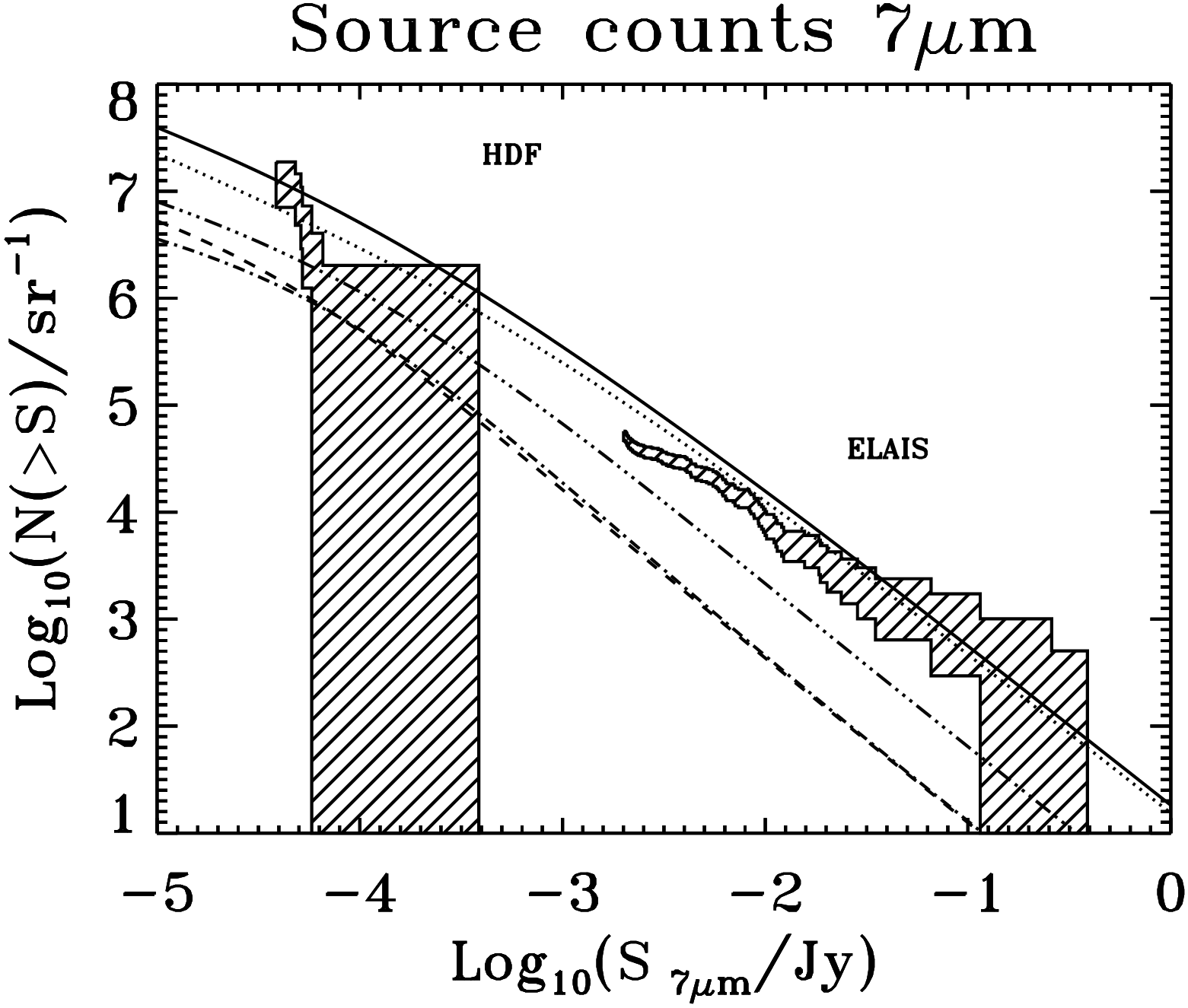}
  \ForceWidth{3in}
  \hSlide{0cm}
  \BoxedEPSF{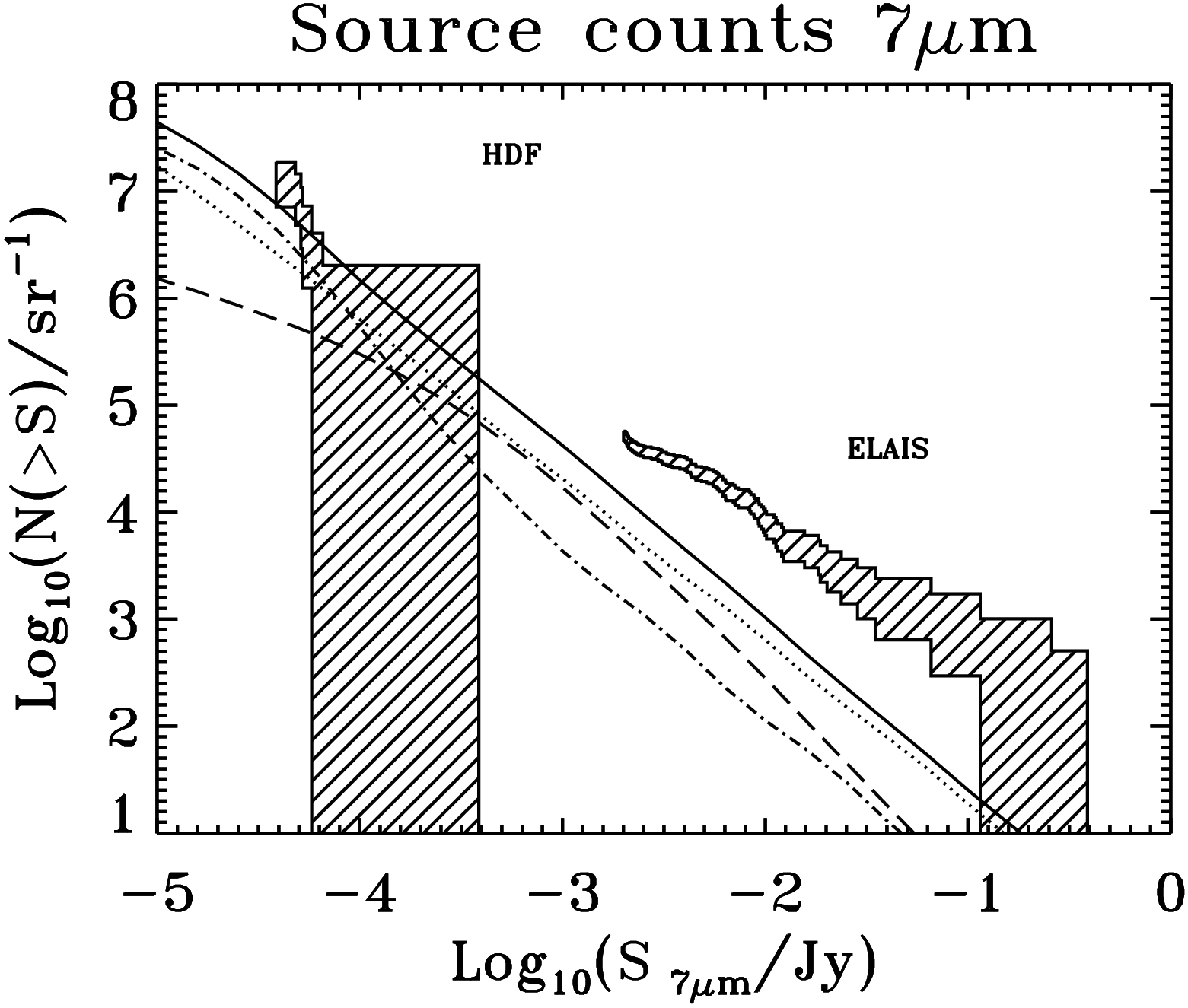}
  \vspace*{0.5cm}

\caption{\label{fig:counts7}
Preliminary integral ELAIS extragalactic source counts at $6.7\mu$m,
using the flux calibration discussed in the text. 
Flux densities are quoted in Jy. 
Left panel shows the Pearson \& 
Rowan-Robinson model overplotted, and right panel shows the 
Franceschini et al. (1997) model. 
Symbols as in the respective panels of figure
\ref{fig:counts}. 
}
\end{figure*}


\begin{figure*}
\centering


  \vspace*{-2cm}
  \ForceWidth{3in}
  \hSlide{-1cm}
  \BoxedEPSF{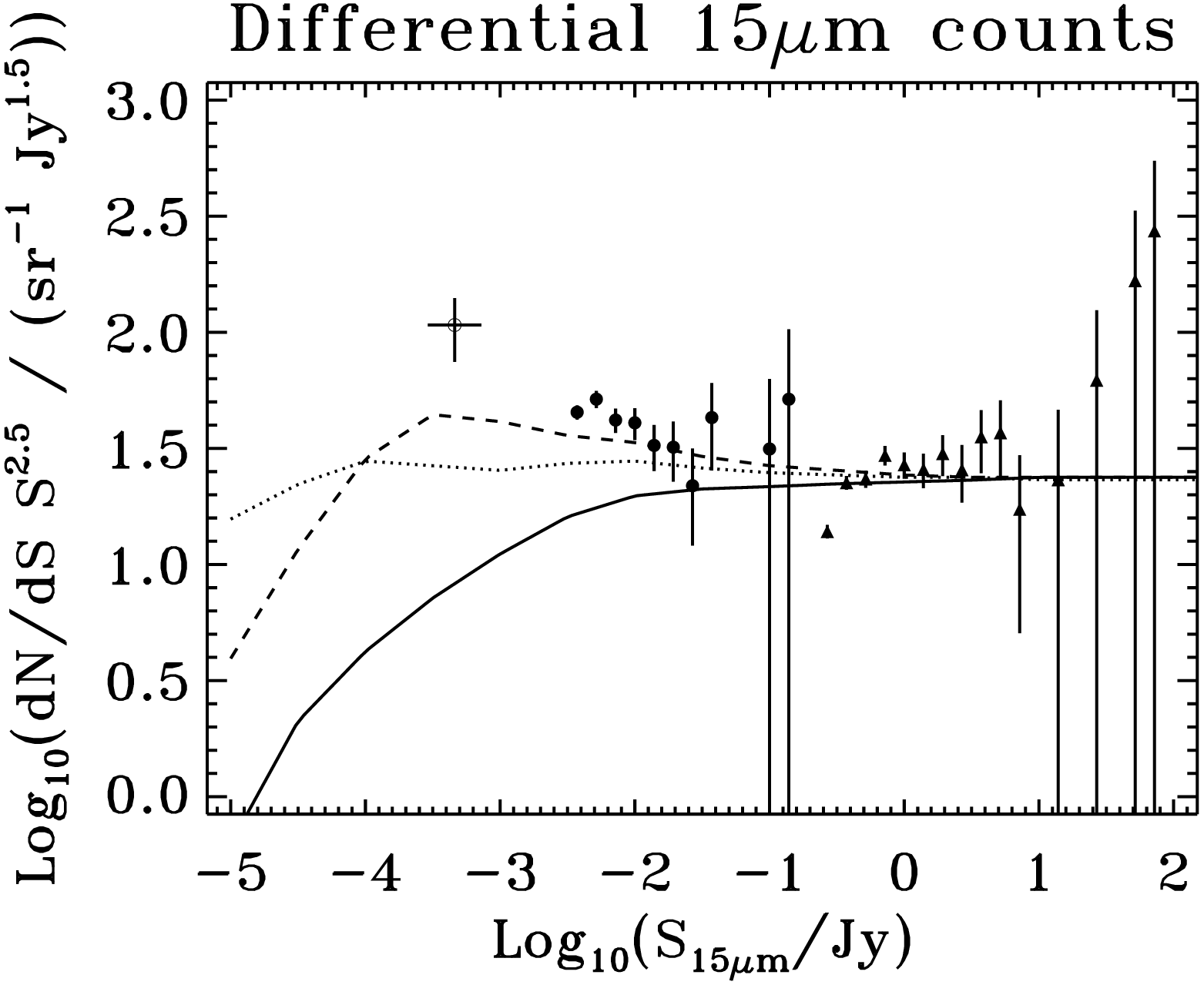}
  \ForceWidth{3in}
  \hSlide{0cm}
  \BoxedEPSF{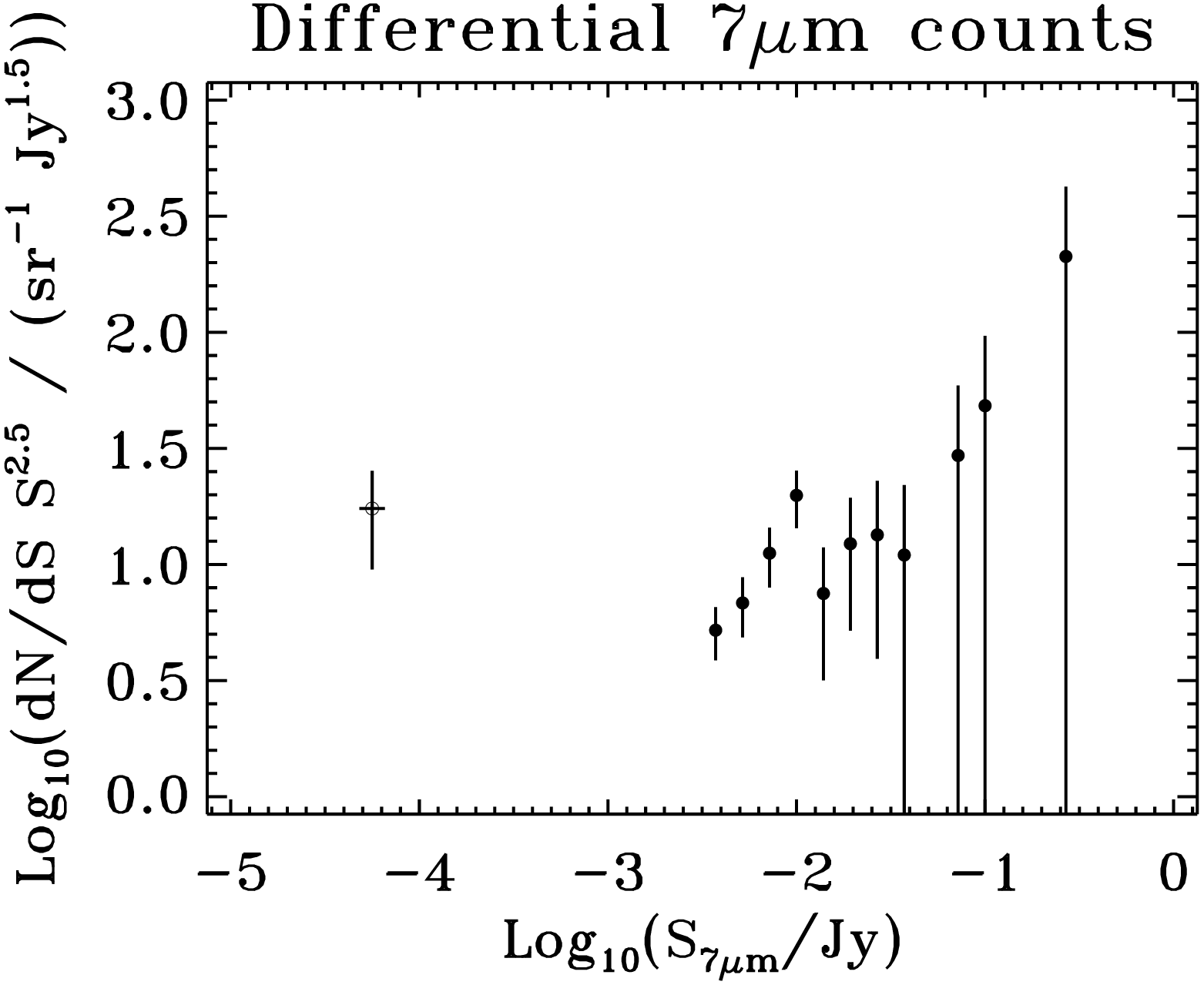}
  \vspace*{0.5cm}

\caption{\label{fig:differentialcounts}
Left panel shows Euclidean-normalised differential source counts at
$15\mu$m. Flux 
densities are quoted in 
Jy. Triangles show the Rush et al. (1993) sample, and filled circles
show ELAIS. The 
open circle is taken from 
the ISO-HDF sample
(Oliver et al. 1997). The filled line is the no-evolution model of Xu
et al. (1998) with mid-IR spectral features incorporated. The dotted
and dashed lines respectively show the Xu et al. (1998) predictions
for $(1+z)^3$ pure luminosity evolution and $(1+z)^4$ pure density
evolution, both including mid-IR spectral features. All three Xu et
al. (1998) models have been renormalised by $\times 1.8$ to match the
IRAS counts. Horizontal error bars (where shown) indicate the width
of the flux bins for the fainter samples, 
and 
vertical error bars the
$\pm 1 \sigma$ Poisson 
errors on the counts. 
The right panel shows Euclidean-normalised differential source counts
at $6.7\mu$m. Flux 
densities are quoted in 
Jy. Filled circles show ELAIS, and the open circle shows the ISO-HDF
sources (Oliver et al. 1997) with optical IDs. 
}
\end{figure*}


\subsection{Optical identifications}\label{sec:optid}
In this section we summarise the optical identification algorithm
used for the Preliminary Analysis catalogue, for APM stars and galaxies (McMahon \&
Irwin 1992). 
We adopted the likelihood ratio procedure of Sutherland \& Saunders
(1992) to associate
our Preliminary Analysis sources with known objects. The surface density of catalogue
objects as a function of magnitude is incorporated into the
likelihood of each identification of the Preliminary Analysis sources with the
catalogue objects. Following Mann et al. (1997), we define the
likelihood ratio to be the ratio of the
probability of detecting a 
genuine counterpart to the source with the position and flux of the
catalogue object, to the probability of such an association occurring
by chance given the positional errors. For a catalogue surface density
$n(f)$ (where $f$ is the flux), a positional uncertainty
$\epsilon(x,y)$ (where $x$ and $y$ are e.g. Cartesian coordinates) and
an a priori flux distribution of IDs given by $q(f)$, the likelihood
ratio is given by
\begin{equation}
LR(f,x,y) \propto \frac{q(f)\epsilon(x,y)}{n(f)}
\end{equation}
Using this
expression and, assuming $q(f)$ to be constant as a function of optical
magnitude, we found, for each Preliminary Analysis source, the object in the APM catalogue
giving the highest likelihood ratio.
We normalised the likelihood ratios by finding the maximum likelihood-ratio
associations around
random source positions, to yield probabilities for each Preliminary Analysis
identification, and accepted identifications with random probabilities 
of less than $0.3$. 
Slight errors in the lens positioning lead to systematic astrometric
shifts in each raster, of order a few arcseconds. To correct for this,
the high-likelihood identifications in each raster were used to
determine any systematic astrometric offset. The identifications were
subsequently rederived. There were not enough reliable optical
identifications to obtain a robust estimate of the lens offset in the
smaller ELAIS areas, so the analysis was restricted to the main ELAIS
areas of N1, N2, N3 and S1. Further discussion of the identifications
is deferred to later papers in this series. 

\subsection{Source counts}\label{sec:counts}
Using 
the completeness and reliability from section \ref{sec:comp_rel},
we can extract the extragalactic source counts at
$6.7\mu$m and $15\mu$m. 
For the purposes of the counts we included all sources accepted by (at
least) two observers.  
From this list, we exclude $15\mu$m stellar identifications brighter
than approximately $B=16.5$ (assuming a monotonic stellar plate
saturation correction converting $B=18$ galaxy magnitudes to
$B\stackrel{<}{_\sim}16.5$ stars), but include fainter stellar IDs
since these are expected to be predominantly AGN. All stellar IDs at
$6.7\mu$m were eliminated from the extragalactic source counts.
It is highly
unlikely for stars with $B>16.5$
to be detected
at $15\mu$m to these limits ($B=16.5$ is equivalent
to $S_{\rm B}=1$ mJy; see also Crockett et al. 1999
and Crockett 1999), with the exception of rare dust-shell stars. 
Note also that all the $15\mu$m ($6.7\mu$m) sources accepted by only
one observer are fainter than $2.2$mJy ($2.5$mJy), so we can be
highly confident of the reliability of sources brighter than these
limits.

The segregation of extragalactic from stellar counts is robust 
at $15\mu$m, but the large fraction of stellar IDs at $6.7\mu$m make
it possible that some faint stars have been included in the
extragalactic counts at this wavelength: of the $794$ 
doubly-accepted sources at $6.7\mu$m, only $79$ have stellar APM
classifications faint enough to be accepted in the extragalactic
counts. We eyeballed the DSS images of every optical
identification of the $6.7\mu$m and $15\mu$m sources. We found the
by-eye classifications to agree with the APM in nearly all cases. The
resultant extragalactic counts are virtually indistinguishable from 
the automated segregations at both $6.7\mu$m and $15\mu$m. 

The extragalactic source counts are plotted in figure
\ref{fig:counts}, 
using the by-eye star-galaxy separation and
the stellar flux calibration. 
The $15\mu$m Lockman Hole ISOCAM survey data will be discussed in
Elbaz et al. 1999. The counts from this survey at around the
$\sim5$mJy level, which overlap with the ELAIS counts, appear
significantly lower than those of ELAIS. However this is entirely
attributable to the $\sim\times2$ differences in assumed 
flux calibration. The counts are in excellent agreement in 
instrumental units or in mJy with the same flux calibration
assumption; alternatively, a $\sim30\%$ 
reduction of the ELAIS Preliminary Analysis $15\mu$m flux calibration
factor would also bring the 
counts into formal $1\sigma$ agreement while remaining consistent with 
figure \ref{fig:fluxcal}. However this may require a commensurate
reduction in the IRAS counts. Such a reduction has been argued to be
necessary by Elbaz et al. (1999) in order to account for
large-scale-structure effects in the Rush et al. sample.

Also plotted 
are  
the $12\mu$m source counts 
calculated by Oliver et al. (1997) from the Rush et al. (1993) sample,
using the QMW IRAS Galaxy Mask (Rowan-Robinson et al. 1991), and
transposed to $15\mu$m assuming a population mix matching the Pearson
\& Rowan-Robinson (1996) predictions. The faint source
counts from the ISO-HDF North survey (Oliver et al. 1997) are also
shown in the counts figures. 
The ELAIS 
extragalactic Preliminary Analysis counts at $15\mu$m  
are consistent with an interpolation between the ISO-HDF North 
and Rush et
al. (1993) data sets, and at $6.7\mu$m with reasonable extrapolations
from the ISO-HDF North. 
The source density at $15\mu$m is also in good agreement with that
obtained at $12\mu$m by Clements et al. (1999), though the differing
K-corrections make it not immediately clear that the counts must
necessarily agree (e.g. Xu et al. 1998, Serjeant 1999). 
Note that the $6.7\mu$m counts have significant
photometric errors independent of flux, due to the undersampling of
the PSF 
(figure \ref{fig:fluxcal}). 
For a
constant source count slope the shape of the counts is unaffected by
flux-independent errors, so we can regard the $6.7\mu$m counts as
subject to a potential systematic error in the form of a horizontal
shift. Such a systematic is less than or of order a factor $2$ in flux. 
We overplot the model predictions from Pearson \& Rowan-Robinson
(1996) in figures \ref{fig:counts} and \ref{fig:counts7}, as well as
the 
model predictions of Franceschini et al. (1994, 1997).
Also overplotted are the 
Guiderdoni et
al. (1997) $15\mu$m 
model counts 
and 
the evolving models from Xu et al. (1998), with and without mid-IR
spectral features.
Note that all the Xu et al. models have been renormalised to match the 
IRAS counts. 
Figure \ref{fig:counts} also 
compares the observations to a variety of non-evolving models. Apart
from the renormalisation to the IRAS Rush et al. (1993) $\times$ QIGC counts,
these are the same non-evolving models as 
discussed in Elbaz et al. 1998. 

\section{Discussion}\label{sec:discussion}
The experimental agreement with the evolving models of Pearson \&
Rowan-Robinson (1996), Franceschini et al. (1994), Guiderdoni et
al. (1997) and Xu et al. (1998) at $15\mu$m over seven
orders of magnitude in flux density is striking. 
The starbursts in the Pearson \& Rowan-Robinson (1996) models are
normalised to the 
$60\mu$m IRAS counts, but not explicitly to the $12\mu$m counts. 
The slight overprediction of the IRAS
counts is also present in the ELAIS counts, until a slight upturn at
around $10$mJy (which is not an effect of incompleteness or low
reliability) departing from the Euclidean slope brings the 
data into closer agreement with the model. 
Of the four Xu et al. (1998) models, the $(1+z)^3$ luminosity
evolution models have the stronger upturn, reproducing the observed
counts slightly more well than the alternative $(1+z)^4$ density
evolution. This is clearer still in the $15\mu$m differential counts
(figure \ref{fig:differentialcounts}), where luminosity
evolution is shown to make a much better fit 
to the ELAIS counts, though an even larger excess is suggested by the
ISO-HDF counts. 
Note
that we renormalised the Xu et al. (1998) predictions by $\times1.8$ to
match the IRAS counts. 


The source count models present rather different predictions at
$6.7\mu$m. 
The Pearson \& Rowan-Robinson (1996) model has only a slight
overestimate, accountable for instance to the flux calibration 
uncertainties. However, the underprediction in the Franceschini et al.
(1997)
model is over an order of magnitude, probably too large to be an
artefact of 
our albeit uncertain flux calibration at this wavelength. 
This discrepancy is most likely to be due to
deficiencies in the assumed spectral energy distributions, which are
not well-constrained in this wavelength range (e.g. Serjeant 1999),
rather than due to incorrect assumptions about the evolution or
population mixes. 

The loss of the Wide Field Infrared Explorer (WIRE) satellite was a
serious blow to infrared extragalactic astronomy. In the hope or 
expectation of a new WIRE mission, we can compare our source counts
with the expectations of the WIRE team. 
WIRE was to survey at least $170$ square degrees to a limiting
$12\mu$m sensitivity of $1.9$mJy, and smaller areas to deeper limits. 
Our source counts imply a surface density of approximately
$100$ galaxies per square degree to the projected WIRE $12\mu$m bright
survey limit, using the Pearson \& Rowan-Robinson or Franceschini et
al. $15\mu$m  source count models to
extrapolate. This is 
larger than projected source density of the WIRE team
(q.v. figure \ref{fig:counts}, bottom-right panel, and 
\ref{fig:differentialcounts} left panel, where
the WIRE predictions were renormalised upwards by a factor of $\sim2$
to match e.g. the observed IRAS and ELAIS counts). Although
encouraging for the numbers of sources, it suggests that surveys much
deeper than this 
will very rapidly be confusion limited. 
This is in good agreement with the observations of 
Oliver et al. (1997). 
The composite counts from the various $15\mu$m surveys also imply that 
confusion will be irrelevant
e.g. for NGST, but will be 
significant for the MIPS imager on SIRTF even in short
($\sim 2000$ second) exposures. The gains in detector size and
sensitivity in future missions will offer very large improvements in
wide-field survey efficiency: for example, 
a survey of a similar
depth and areal coverage to ELAIS at $15\mu$m will be possible with
MIPS on SIRTF in only $\sim 10$ hours.

%

The upturn in the counts at faint fluxes continues in the 
Hubble Deep 
Field ISO counts, and is too large to be attributable to
starburst K-corrections (e.g. Xu et al. 1998, Elbaz et al. 1998a,b). 
This excess is above any reasonable no-evolution predictions
(figure \ref{fig:counts}), 
which all have shallower slopes than the evolving models
independent of K-correction and world models. 
This implies that 
the faintest sources in ELAIS are sampling a significantly
cosmologically evolving mid-IR population. Given the strong evolution
in both the comoving star formation rate claimed from optical-UV
observations and in the quasar comoving number density, as well as the 
large fraction of AGN and starbursts expected in ELAIS from the counts 
models, we can
reasonably expect
that optical spectroscopy of ELAIS and fainter samples 
will have a major
impact on the study of dust-shrouded star formation and quasar
activity and their evolution. The prediction 
from the Pearson \& Rowan-Robinson (1996) source count models is for 
of order a few objects as (intrinsically) luminous as IRAS F10214+4724
out to $z=4$ over the
entire ELAIS areas. The MIR luminosity function and luminosity density 
can be used to derive a comoving star formation rate without the
difficult problems of extinction corrections that affect optical-UV
estimates, and (for the ISOCAM LW3 filter) is reasonably free of
K-correction uncertainties (Xu et al. 1998, Serjeant 1999). 
The ELAIS survey is also highly sensitive to obscured
quasars and will be an exceptional resource for active galaxy
unification models. The ELAIS limits can also provide important
constraints for sources detected at other wavelengths: for example,
the $15\mu$m, $90\mu$m and $175\mu$m limits at $850\mu$m source
positions can provide robust redshift constraints (e.g. Hughes et
al. 1998).

\section{Conclusions}\label{sec:conclusions}
The extragalactic source counts agree extremely well with all evolving 
model
predictions (Franceschini et al. 1994, Pearson \& Rowan-Robinson 1996,
Guiderdoni et al. 1997, Xu et al. 1998)
over seven orders of magnitude in $15\mu$m flux 
density. The Pearson \& Rowan-Robinson (1996) models can broadly
reproduce the $6.7\mu$m extragalactic counts, but the observations are 
in excess of the Franceschini et al. (1997) predicted counts at this
wavelength using our preliminary $6.7\mu$m flux calibration. All 
no-evolution models are clearly
excluded, and imply a 
cosmologically evolving population
of obscured starbursts and/or active galaxies 
dominates below $\sim10$ mJy at $15\mu$m, independent of K-correction
assumptions. 
Source confusion appears to have been underestimated in the
WIRE and SIRTF missions, but will not impact significantly on the NGST. 

\section*{Acknowledgements}
We would like to thank Dave Alexander and Ruth Carballo for helpful
comments and proofreading of this paper. 
This paper is based on observations with ISO, an ESA project with
instruments funded by ESA 
member states (especially the PI countries: France, Germany, the
Netherlands and the United 
Kingdom) and with participation of ISAS and NASA. 
The ISOCAM data presented in this paper was analysed using ``CIA'', 
a joint development by the ESA Astrophysics Division and the ISOCAM     
Consortium. The ISOCAM Consortium is led by the ISOCAM PI, C. Cesarsky,  
Direction des Sciences de la Matiere, C.E.A., France.
This work was
supported by PPARC (grant 
number GR/K98728) and by the EC TMR Network programme
(FMRX-CT96-0068).

\end{document}